\documentclass[aps,pra,twocolumn,amsmath,amssymb,superscriptaddress,showpacs, 10pt]{revtex4-1}

\usepackage[utf8]{inputenc}
\usepackage{amsfonts}
\usepackage{amsmath}
\usepackage{amssymb}
\usepackage{amsthm}
\usepackage{fontenc}
\usepackage{graphicx}
\usepackage{xcolor}
\usepackage{textcomp}
\usepackage{epstopdf}
\usepackage{braket}
\usepackage{mathtools}
\usepackage{amsmath}
\usepackage{dcolumn}
\usepackage{multirow}

\begin{document}
\newcommand{\abs}[1]{\lvert#1\rvert}
\title{  
Gap engineering in strained fold-like armchair graphene nanoribbons}
\author{V. Torres}
\affiliation{Instituto de F\' isica, Universidade Federal Fluminense, Niter\' oi, Av.~Litor\^anea sn 24210-340, RJ-Brazil}
\author{C. Le\' on}
\affiliation{Instituto de F\' isica, Universidade Federal Fluminense, Niter\' oi, Av.~Litor\^anea sn 24210-340, RJ-Brazil}
\author{D. Faria}
\affiliation{Instituto Polit\' ecnico, Universidade do Estado do Rio de Janeiro, Nova Friburgo, 28625-570, RJ-Brazil}
\author{A. Latgé}
\affiliation{Instituto de F\' isica, Universidade Federal Fluminense, Niter\' oi, Av.~Litor\^anea sn 24210-340, RJ-Brazil}

\date{\today}

\begin{abstract}

Strain fold-like deformations on armchair graphene nanoribbons (AGNRs) can be 
properly engineered in experimental setups, and could lead to a new controlling tool for gaps and transport properties. Here, we analyze the electronic properties of folded AGNRs relating the electronic responses and the mechanical deformation. An important and universal parameter for the gap engineering is the ribbon percent width variation, i.e., the difference between the deformed and undeformed ribbon widths. AGNRs bandgap can be tuned mechanically in a well defined bounded range of energy values, eventually leading to a metallic system. This characteristic provides a new controllable degree of freedom that allows manipulation of electronic currents. We show that the numerical results are analytically predicted by solving the Dirac equation for the strained system. 

\end{abstract}

\maketitle

\section{Introduction}

Graphene nanoribbons are known as excellent counterpart of graphene due to the possibility of exhibiting accessible and modeled energy gaps\cite{Castro2009,Wakabayashy2010, Ritter2008}.  In particular, ribbons with armchair edge belong to different semiconducting families, with a gap size depending essentially on the nanoribbon width. Zigzag ribbons, otherwise, show metallic behavior due to the presence of edge localized states at the Fermi energy\cite{Brey2006}. The presence of roughness, mainly at the edges of the ribbons, can be responsible for important transport suppressions and formation of conductance gaps, leading to some restrictions for the use of graphene in ballistic devices\cite{Mucciolo2009}. Differently from what could be naively expected, some particular deformations can however strength transport properties as it happens in the case of strained-fold graphene systems. 

 When graphene samples are deposited on $SiO_2$ or hexagonal $BN$ substrates, deformations such bubbles and wrinkles may accidentally arise mainly due to the difference in the thermal expansion coefficients of the constituent materials\cite{Levy2010,Lim2015}. Deformations can then be produced, leading to the formation of homogeneous pseudo-magnetic fields\cite{Downs2016}. Other routes are based on patterning not the graphene itself but the supporting substrates that induce different strain profiles \cite{Pereirax2009}.  Physical scenarios have been explored where strain induced pseudo-Landau levels are observed \cite{Klimov2012,Levy2010,Li2015,Guinea2010}. In addition, controlled creation of periodic ripples in suspended graphene sheets may also be achieved by thermally generated strains\cite{Bao2009}. Alternatively, strained folds can be generated by pressuring a gas inside a sealed container with a slit covered by graphene \cite{McEuen} or by the presence of a gate voltage, below or on top of graphene on an extended trench to induce the deformation \cite{Fogler}. A large quantity of theoretical works have discussed the coupling between mechanical deformations and electronic responses for particular deformations in graphene \cite{Schneider2015,Wakker2011, Pereira2009, Moldovan2013, Ramon2014, Daiara2013,Zenan2013,Ramon2016,Polini2011}. Experimentally, laser ablation, STM and AFM have been used for manipulation and detection of these deformations\cite{Lim2015,Klimov2012,Xu2012}.

Recently the electronic transport along fold-like deformed areas have been explored in zigzag nanoribbons\cite{Ramon2016, Polini2011,Prada2010}. When a gaussian-like deformation parallel to the zigzag direction is considered,  strain-induced pseudomagnetic fields are formed exhibiting stripped spatial distribution. It was shown that folded zigzag GNRs behave as natural waveguides for electronic transport and provide a splitting of an incident current, which results in valley-polarized currents moving along different parts of the structure\cite{Ramon2016}.  The spatial separation of valley currents is preserved when disorder at the edge are included in the model, giving rise to novel quasi-ballistic transport characteristics. On the other hand, folded armchair nanoribbons do not generate pseudomagnetic fields like the zigzag GNRs. 
\begin{center}
\begin{figure}[hbt]
\includegraphics[width=1.0\columnwidth]{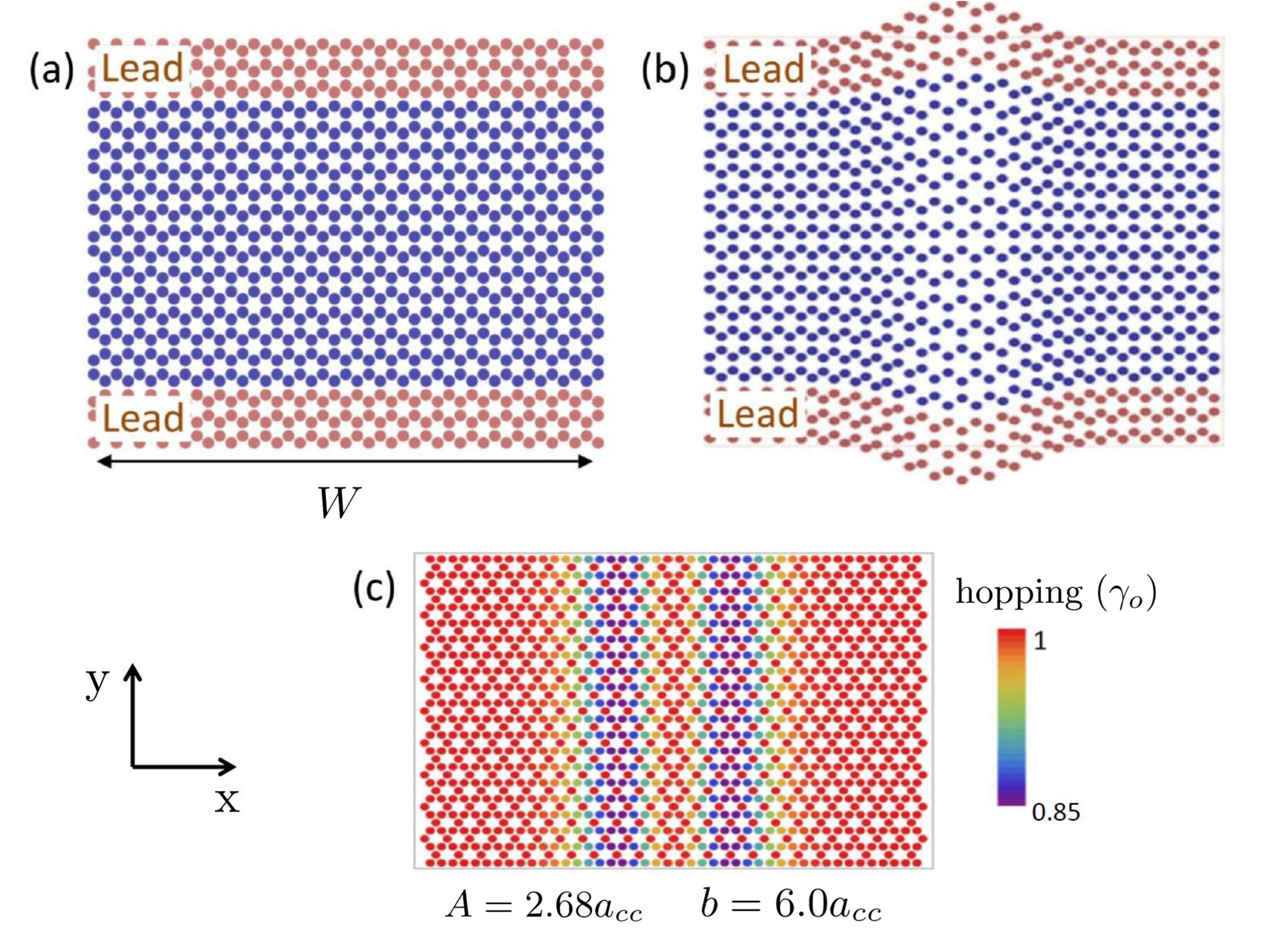}
\caption{(Color online) Schematic view of an (a) unstrained and a (b) strained-fold  AGNR of width $W$ given by 45 carbon atoms. (c) Colored plot of the hopping energy distribution for a strained ribbon with Gaussian parameters  $A=2.68a_{cc}$ and $b=6.0a_{cc}$.} 
 \label{fig:device}  
\end{figure}
\end{center}
As the fold axis changes from the zigzag to the armchair direction, with the fold parallel to the nanoribbon edge, the pseudomagnetic changes from a maximum value to zero field. As such, the armchair GNRs do not support the previous mentioned electronic waveguides. Then the natural question that rises is how the electronic transport is affected in strained fold armchair GNRs.

The effects of in-plane strain, uniaxial and shear, on graphene nanoribbon band structures have been discussed\cite{Lu2010,Sun2008}.  The induced vector potential changes the distance between the Dirac valleys affecting the gap at the $\Gamma$ point. The energy gap of armchair ribbons with N carbon atoms along the ribbons width, N-AGNRs, are found to vary linearly or periodically with uniaxial strain for weak and large intensities, respectively. Interestingly, the two semiconducting families exhibit different behavior as a function of the uniaxial strain; while for family $3m+1$ the gap energy increases with the deformation, the gap decreases as a function of increasing strain for the $3m$ family.  

In this paper, similar results are found for fold-like out-plane deformations in armchair GNRs. However, we show that the energy gap can be drastically modulated by changing the deformation parameters, amplitude and extension of the fold deformation. Moreover, we discuss that the relative length of the deformed structure plays a key role determining the energy gap and its maximum values. Simple tight binding calculations using real-space renormalization techniques for obtaining the system Green's functions are compared with analytical predictions done by solving the Dirac equation. We also discuss the possibility of a semiconducting AGNR to turn on a metallic ribbon. In addition we analyze the difference and similarities of the band structure results and of the probability density distributions between a natural metallic nanoribbon and an original semiconducting ribbon that becomes metallic. 
 
\section{Structures and model}

The system is composed of a central conductor connected by top and bottom leads, all the three parts being perfect armchair graphene nanoribbons. The ribbon width is $W=(N+1)\sqrt{3} a_{cc}/2$, with $a_{cc}=1.42\,\textrm{\AA}$ being the interatomic distance and $N$ the number of carbon atoms along the nanoribbon width.  A schematic view of the armchair ribbon is shown in Fig.\,\ref{fig:device} (a). A single $\pi$-band tight binding Hamiltonian is used to describe the system, given by

\begin{equation} 
H =\sum_{\substack {<i,j>}} \gamma_{i,j} c_{l}^\dagger  c_{m}  + h.c.\,\,\, 
\end{equation}
with $c_i^\dagger$ ($c_i$)  being the  creation  (destruction) operator for an electron in site $i$ and $\gamma_{i,j}$ the nearest-neighbor hopping, that in the case of pristine graphene is $\gamma_o \approx 2.75 eV$. In our model, the fold-like mechanical deformation is described using the linear elasticity theory \cite{Landau1970, Katesnelson2012}, with the strain tensor written in terms of the in- and out-of-plane deformation, $u_{\mu}$ and $h$, respectively\cite{Moldovan2013,Ramon2014},
\begin{equation}
\varepsilon_{\mu \nu}=\frac{1}{2}\left(\partial_\nu u_\mu+\partial_\mu u_\nu+\partial_\mu h \partial_\nu h\right)\,\,.
\end{equation}
Within the microscopic approach, there is a change in the distance between first-neighbor carbon atoms, compared to the interatomic distance of the unstrained system $a_{cc}$, expressed as
\begin{equation}
l_{ij}=\frac{1}{a_{cc}}\left(a_{cc}^{2}+\varepsilon_{xx}x_{ij}^2+\varepsilon_{yy}y_{ij}^2+2\varepsilon_{xy}x_{ij}y_{ij}\right)\,\,,
\end{equation}
where $x_{ij}$ and $y_{ij}$ are the horizontal and vertical projections of the lattice vectors in the undeformed ribbon, respectively. This change in distance is included in the modified nearest-neighbor hopping energy $\gamma_{ij} = \gamma_{o}\exp(-\beta ( \,\, l_{ij}/a_{cc}  - 1))$, with $\beta$ being the electromechanical coupling strength, $\beta=\left|\partial\log t_{o}/\partial\log a\right| \simeq 3$. 

In the continuum description, the effect of the deformation usually appears as an inhomogeneous pseudo-gauge field \cite{Castro2014,Sloan2013,Settnes2016,Daiara2013}. The hopping modifications give origin to gauge fields in the Dirac equation \cite{Ando}, with the pseudo vector potential  written in terms of the strain tensor elements,
\begin{equation}
(A_x,A_y)=\frac{\beta\hbar v_f}{2a_{cc}}(\varepsilon_{xx}-\varepsilon_{yy};-2\varepsilon_{xy})\,\,,
\end{equation}
and raising a pseudomagnetic field $\vec{B} = \nabla\times\vec{A}(\vec{r})$.  A scalar deformation potential $V=g(\varepsilon_{xx}+\varepsilon_{yy})$ can be also considered in the Hamiltonian, with additional diagonal contributions.  The parameter $g$ describes the coupling to acoustical phonons in graphene systems, and a wide range of values (0 to -20 eV) has been adopted in different calculations\cite{Vozmediano2010},  providing an effective way to take into account the potential screening in graphene.  
 
We consider an extended Gaussian deformation along the whole system, as illustrated in Fig.\,\ref{fig:device}(b) and described by:
\begin{equation}
h(x_{i})=A\exp\left(-\frac{(x_{i}-x_{0})^{2}}{b^{2}}\right)\,\,\, ,
\label{eq:vector potential}
\end{equation}
which runs over the finite size confined direction, $x$. Here, $A$ and $b$ parametrize the Gaussian amplitude and width, respectively, and $x_0=W/2$ defines the position of the ribbon central axis. Fig. \,\ref{fig:device} (b) shows a strained 45-AGNR with $A=2.68 a_{cc}$ and $b=6.0 a_{cc}$.

The new atomic distance for the strained fold deformation, along the zigzag direction, is given by
\begin{equation}
l_{ij}=\frac{1}{a_{cc}}\left(a_{cc}^{2}+\varepsilon_{xx}x_{ij}^2\right)= a_{cc}\left(1+\frac{3 (x-x_0)^ 2 h^2(x)}{2b^4}\right)   \,\,,
\end{equation}
evaluated for $x_{ij}=\sqrt{3}/2a_{cc}$. It is possible to obtain the maximum distance variation between the atoms, that is $\Delta\l_{m}/a_{cc}=(l_{ij}-a_{cc})/a_{cc}=3\alpha/4e$, for $x=\pm b/\sqrt{2}$, where $\alpha=(A/b)^2$ and $e$ is the Euler's number $(e= 2.71828...)$. Then, in what follows we use the variable $\alpha$ to indicate the strain intensity considered in the system. The spatial dependence of the hopping energies $\gamma_{ij}$ for the strained- fold AGNR is shown in Fig.\,\ref{fig:device}(c). The colored diagram maps the hopping energy at the mean distance between atoms $i$ and $j$. For the strain parameters considered ($\alpha=20\%$), the maximum distance variation between neighbor sites is  $\Delta\l_{m}=5.5\%$, while the highest hopping modification is $15\%$.

Differently from the case of a strained fold nanoribbons with zigzag edges, which presents a pseudomagnetic field configuration in the central part of the ribbon \cite{Ramon2016}, a strained fold armchair nanoribbon does not give rise to pseudomagnetic fields although the vector and scalar potential are non null quantities. 
For the extended Gaussian deformation  considered, the vector potential and the scalar potential are given, respectively,  by $\left(A_{x}, A_{y}\right)=\left(\varepsilon_{xx},0\right)$ and  $V(x) = g \varepsilon_{xx}$. It is also possible to understand these electronic properties modifications in terms of the local metric and curvature invariants of the system geometrical distortion, which indicate how much it curves with respect to a nondeformed system\cite{Barraza89}. In particular, the Gaussian curvature $(K)$ is null because the out-of-plane fold deformation varies only on the horizontal direction $x$, while the mean curvature is $H=\varepsilon_{xx}/2(1+\varepsilon_{xx})$, depending locally on the strain at the ribbon. 

It is easy to show that due to the geometric characteristics of the hexagonal lattice, no changes in the interatomic distances are expected for a Gaussian deformation along the armchair transport direction.  As such, AGNR ribbons are not supposed to provide extra conductance channels with localized states along the strained fold-like area as predicted for strained fold zigzag GNRs \cite{Ramon2016}. Nevertheless, the armchair GNRs present interesting variations of the electronic transport that depends on the sizes of the system and on the deformation intensity, as we discuss in the next sections. In particular, the conductance gap size is an important quantity that can be mechanically modulated. 

To calculate the conductance for the AGNR ribbons, we use the Landauer approach within the Green's function formalism\cite{Datta}, written as
\begin{equation}
{G}(\varepsilon) = \frac{2e^2}{h} Tr[\Gamma^{T}(\varepsilon) g^r(\varepsilon) \Gamma^{B}(\varepsilon) g^a(\varepsilon)]\,\,,
\label{GLR}
\end{equation}
where $g^{r(a)} $ is the retarded (advanced) Green's function of the central conductor and $\Gamma^{T(B)}(\varepsilon)=i[\sum^{r}_{T(B))}(\varepsilon) - \sum^{a}_{T(B))}(\varepsilon)]$ is written in terms of the top (bottom) lead -energy $\Sigma^{a,r}_{T(B) \sigma}$. 
\begin{center}
\begin{figure}[h]
\includegraphics[width=0.85\columnwidth]{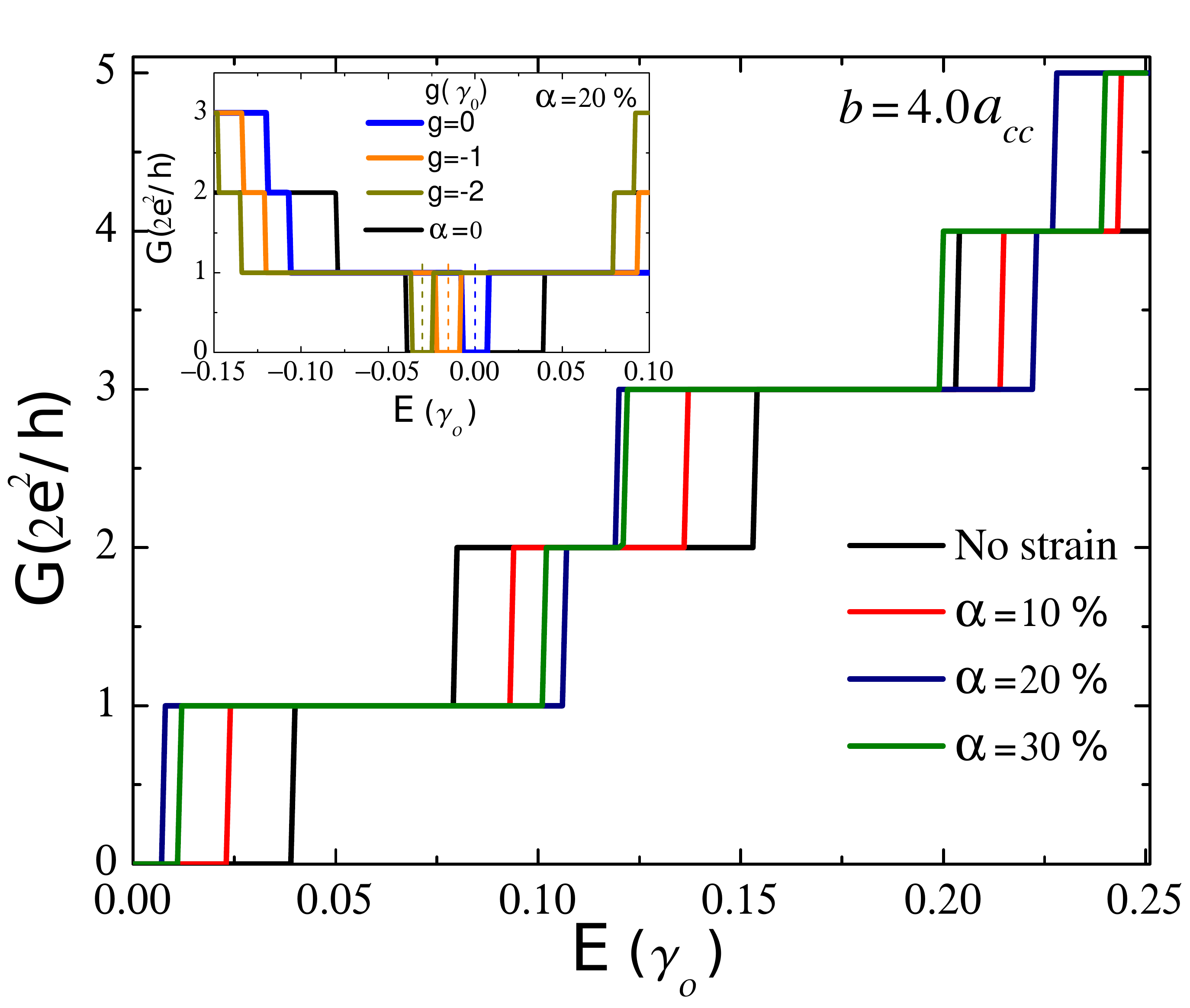}
\caption{(Color online) Conductance as a function of energy for a 45-AGNR with different strained fold amplitudes and fixed width $b=4.0\,a_{cc}$, and null scalar potential. Inset: Conductance  for different values of the coupling parameter $g$ and $\alpha=20\%$.  The colored dashed lines mark the new Fermi energy positions corresponding to the $g$ values.  \label{fig:conductance} }
\end{figure} 
\end{center}

The Green's functions of the leads are calculated using recursive methods, largely explored in different carbon systems \cite{Ritter2008,Latge2000,Caio2013}, while for the central system, circular real-renormalization procedures\cite{Thorgisson2014} are employed. As expected, due to the translation symmetry on the y-direction, the conductances of such fold-deformed ribbons are still marked by a sequence of plateaux as in the case of pristine graphene nanoribbons, but with energy shifts determined by the strain parameters of the theoretical model, as seen in Fig. \ref{fig:conductance}. The results are obtained considering a null scalar potential. Calculating the corresponding conductance we obtain the transport energy gap of each strain configuration. In the inset of Fig. 2 we present conductance results for a particular strain configuration ($\alpha=20\%$) taking into account now the scalar deformation potential $V$ (spatial dependent), that changes the onsite energies of the tight binding Hamiltonian. For the $g$ values considered, we note a shifted of the gap position and also of the corresponding Fermi energies. However,  the size of the conductance gaps, and the semiconducting nature of the AGNR are essentially not altered and then, in what follows, we have neglected this potential contribution. 

\section{Results}

\subsection{Gap dependence: numerical results}

\begin{center}
\begin{figure}[h]
\includegraphics[width=1\columnwidth]{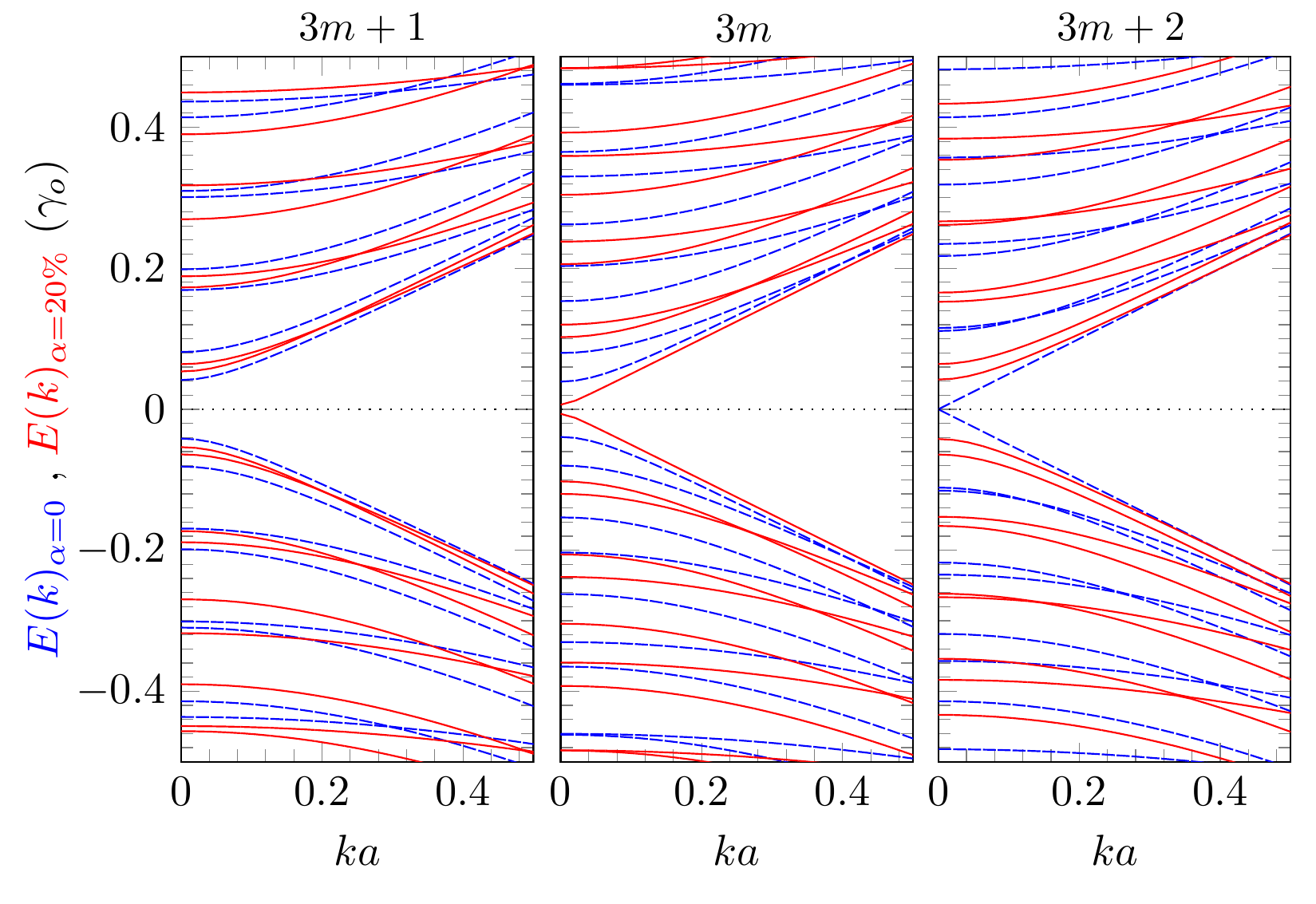}
\caption{(Color online) Electronic band structures of undeformed AGNRs (blue dashed curves) and strained N-AGNRs with $\alpha=20\%$ (red curves), for  $N=43$, $45$ and $47$ atoms along the ribbon width (left, center and right panel, respectively). Parameters: $A=1.34\,a_{cc}$ and $b=3.0\,a_{cc}$.    
\label{fig:FigBandas} }
\end{figure} 
\end{center}

 We start by focusing on the changes of the gap size according with the geometrical parameter of the ribbon deformation.  As the fold deformation is also considered in the leads, the translation symmetry along the nanoribbon fold-axis is preserved, and it is possible to obtain the electronic band structure of the infinite strained ribbon. Results for the electronic structure for the N-AGNR families, $N=3m+1, 3m$, and $3m+2$, under the same deformation are presented in Fig.\,\ref{fig:FigBandas}.  An additional evidence of gap modulation can be inferred by comparing the unperturbed system (dashed blue lines) with the strained fold AGNRs (red continuous curves). Another interesting point is the gap size evolution for the different armchair families. For this particular set of $A$ and $b$ parameters, corresponding to a maximum strain of $5.5\%$, the gap size increases for the $3m+1$ and $3m+2$ cases while it decreased for the $3m$ family.

\begin{center}
\begin{figure}[h]
\includegraphics[width=1\columnwidth]{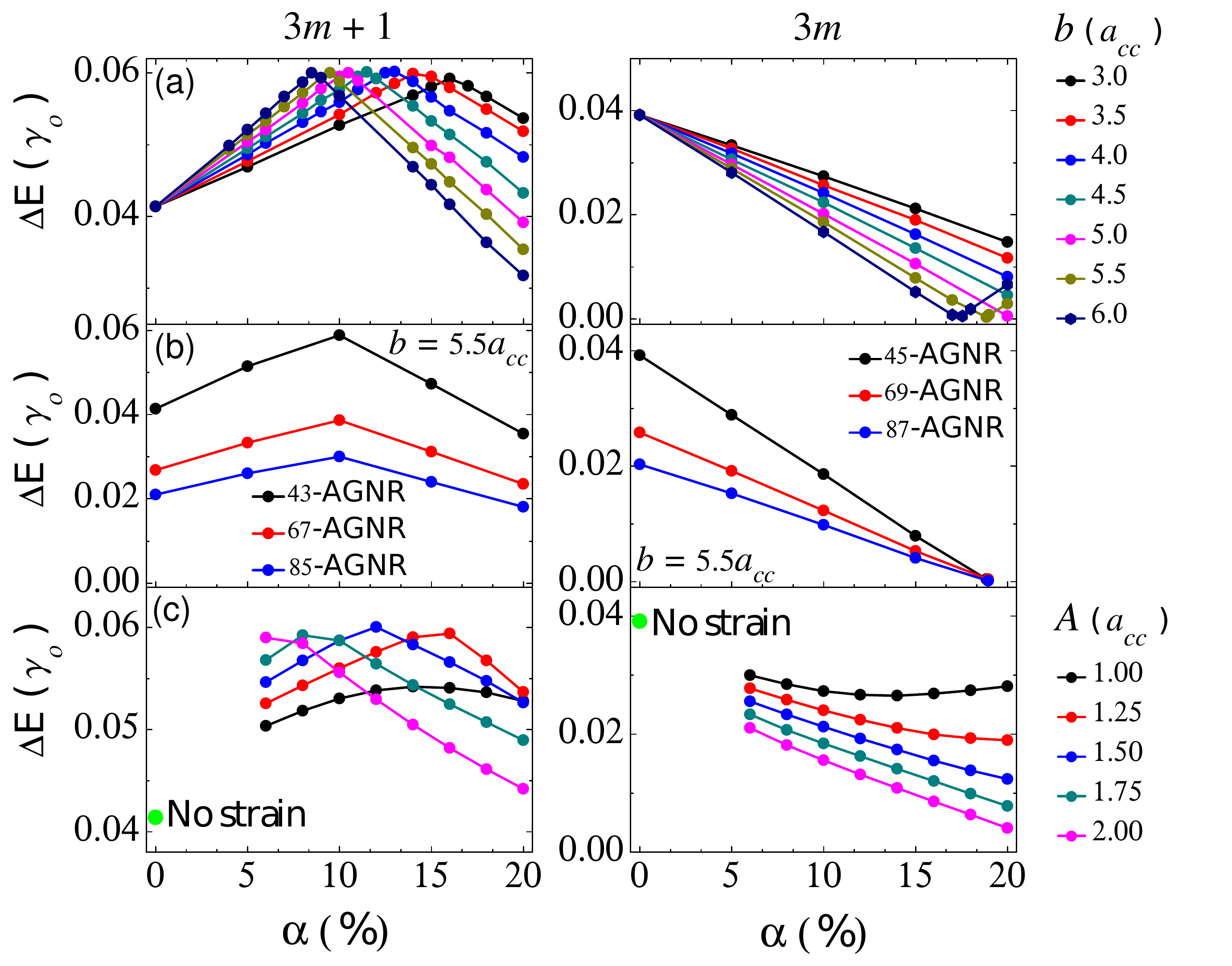}
\caption{(Color online) Evolution of the energy gap (halfwidth) with respect to fold deformation $\alpha$, of two semiconducting N-AGNR families, $N=3m+1$  (left panels) and $3m$ (right panels). (a)  each curve is for a fixed value of standard deviation b, marked in the right legend, for  N=43 (left)  and 45 (right) atoms. (b)  $b=5.5a_{cc}$ and different values of the nanoribbon width. (c) Fixed values of amplitude A, marked in the legend, for  N=43 (left)  and 45 (right) atoms. 
 \label{fig:FigGapVsAlpha} }
\label{gapsx}
\end{figure} 
\end{center}

We present in Fig.\,\ref{fig:FigGapVsAlpha}, the gap size evolution for the two semiconducting N-AGNR families, $N=3m+1$ and $3m$, as a function of the deformation parameter $\alpha$. The results for family $N=3m+1$ are displayed on the left panels, while the results for the family $3m$ are presented on the right panels. Different combinations of the fold-deformation parameters, A and b, are considered.  In Fig.\,\ref{fig:FigGapVsAlpha}(a), the fold width $b$ is constant and the amplitude $A$ varies for different curves. As previously expected, for the $3m+1$ AGNRs the gap increases with $\alpha$ intensity, while the opposite behavior occurs for the $3m$ semiconducting family. This profile is expected to be modified for higher values of strain as the curves change slope signs. 

One can also notice that the gap depends linearly on the $\alpha$ parameter, with the magnitude of slope increasing for higher values of $b$. Another interesting feature is that for the $3m+1$ family the same maximum gap value is achieved at different $\alpha$ values for fixed $b$ width, while for the $3m$ family, a null gap is obtained for different strain intensities and fixed $b$.  As expected, and not shown here, the metallic family $3m+2$ behaves similarly to the $3m+1$ AGNRs for fixed b parameter, increasing the gap energy size from  zero as the deformation is turned on until a maxima value and then going down again. 

In Fig.\,\ref{fig:FigGapVsAlpha}(b), we verify the dependence of the gap size as a function of $\alpha$, with fixed Gaussian width $b$, for different ribbons sizes, for both families. The ribbons considered for $3m+1$ family are:  $N=43, 67$, and $85$ atoms, and for $3m$ family are: $N=45, 69$, and $87$, and $b=5.5\,a_{cc}$. It can easily be seen that the maximum gap for each manoribbon size is different, but they are obtained for the same $\alpha$ parameter for $3m+1$ family. Similarly, for the $3m$ family, the null gap is achieved for a fixed $\alpha$, independently of the nanoribbon width. The dependence of the gap on the parameter $b$, for fixed values of the amplitude $A$ is shown in Fig. \,\ref{fig:FigGapVsAlpha}(c).  The gap values do not vary linear anymore. This shows an additional dependence of the gap on the parameter $b$, than just on $\alpha$. We did not estimate numerically the dependence for small values of $\alpha$, because in this range, $b$ is of the same order than the width of the ribbon, but the curve profiles are expect to varies continuously until they reach the bandgap for unstrained ribbons, marked as a green dot in each panel. For the $3m+1$ family, the gap values are also expected to be bigger for higher deformations, while for the $3m$ family, the gap values are expected to be smaller than the one in the unstrained case. These dependences are further discussed in comparison with analytical results shown in the continuum description section.

\begin{center}
\begin{figure}[h!]
\includegraphics[width=1.0\columnwidth]{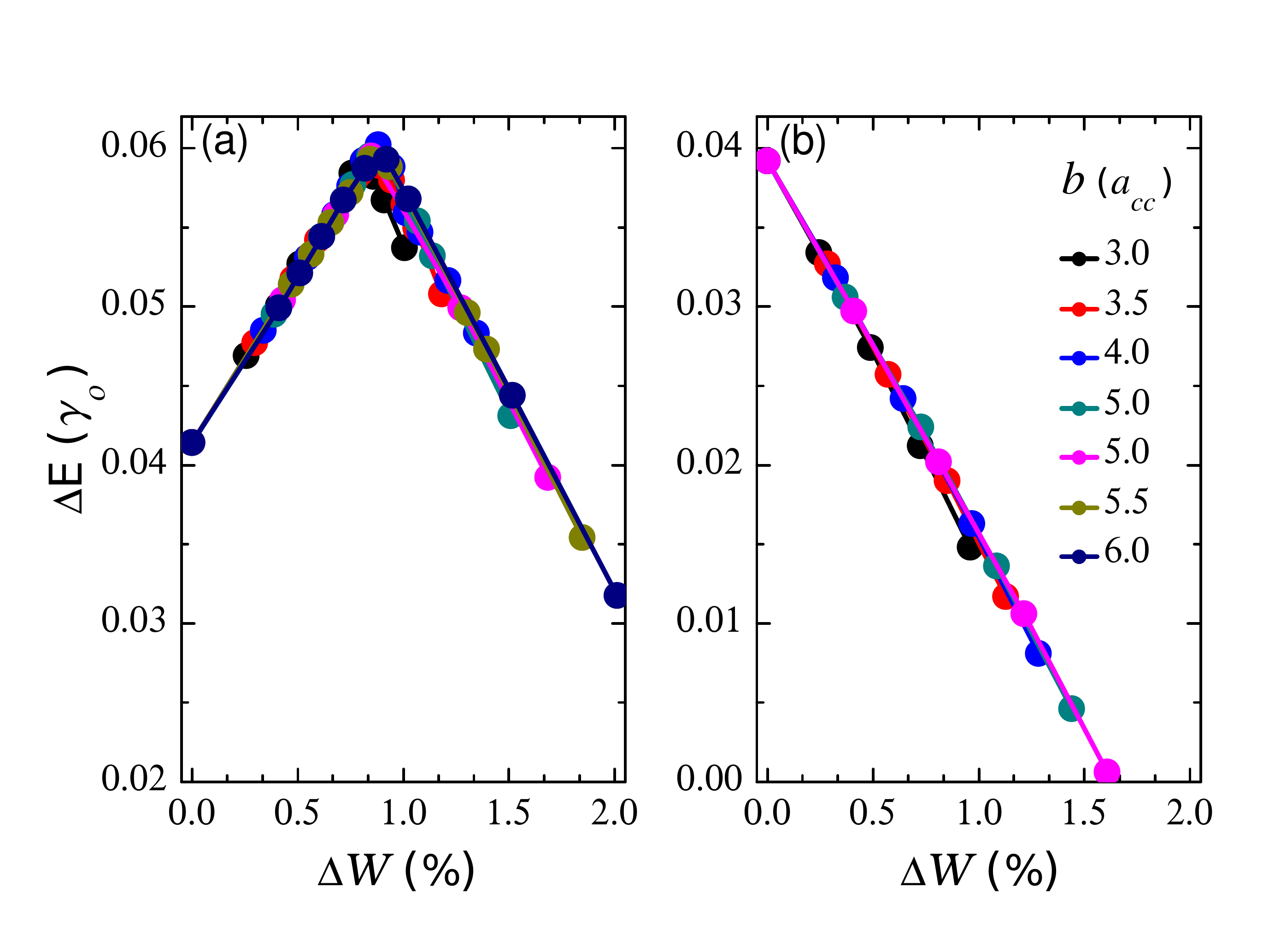}
\caption{(Color online) Evolution of the energy gap (halfwidth) as a function of the percent differential length  $\Delta W(\%)$,  for two semiconducting N-AGNR families, (a) $N=3m+1$ ($N=45$ atoms)  and (b) $N=3m$ ($N=43$ atoms). Fixed values of width $b$ are considered. 
 \label{fig:FigGapVsDeltaL} } 
\end{figure} 
\end{center}

As one of the main results, we show that it is possible to summarize the gap dependence on the fold parameters as a function of the percent width variation of the ribbons after the deformation ($\Delta W/W=\Delta W(\%)$). We define the width variation as $\Delta W= (W_S-W)$, where  $W_S$ is the strained and $W$ is the unstrained ribbon width. In Fig.\,\ref{fig:FigGapVsDeltaL} we show the gap size dependence on the percent width variation of the ribbons. The results show a universal behavior in terms of the deformation parameters for both semiconducting families. This general result indicates that the important parameter to tune the energy gap size, independently of the ribbon width, is the differential length conformed by the out of plane mechanical deformation.

\subsection{Gap dependence: continuum description }

It is possible to understand these results by considering analytically the electron dynamics in strained graphene nanoribbons, which is determined by the Dirac equation in the presence of the gauge field with specific boundary conditions\cite{Pereirax2009}. The gauge field is defined in Eq. \ref{eq:vector potential}. Then, for a given valley, the Dirac equation takes the form $v_f\boldsymbol{\sigma}\cdot\left( \boldsymbol{p}-\boldsymbol{A}/v_f \right)\Psi=\epsilon\Psi $, 
where $\boldsymbol{\sigma}=(\sigma_x,\sigma_y)$, $\Psi=(\Psi_A,\Psi_B)$  is the wave function defined for both A and B sublattices, and energy $\epsilon_n=\pm\hbar v_f\sqrt{k^2_n+k^2_y}$, with $k_n$ being the quantized transversal momentum. The electronic wave function for both $K$ and $K'$ valley is then derived, taking into account the out of plane deformation, given by $h(x)$. Specifically in the present strained-fold case, the gauge field $\boldsymbol{A}$ is oriented along the $x$ direction, and it provides corrections in the mechanical momentum $k_n$. To calculate the transversal momentum, boundary conditions at the edges of the armchair nanoribbon are applied\cite{Brey2006}, and the following relation are obtained

\begin{equation}
\begin{split}
 k_{n}=\frac{n\pi-2\alpha\lambda(0)}{W} -K\,\,,
 \end{split}
\label{eq:transvK}
\end{equation}
with $K={4\pi}/{3\sqrt{3}a_{cc}}$, $n$ labeling the transversal modes and $\lambda (x)$ given by
\begin{equation}
\begin{split}
\lambda(x)= \frac{\beta}{16a_{cc}}
&
\left[\frac{4h^2(x)}{A^2}(x-W/2)\right. 
\\
 &
\left.- b\sqrt{2\pi}Erf\left[ \frac{\sqrt{2}(x-W/2)}{b}\right]  \right],
\end{split}
\label{eq:lambda}
\end{equation}
with the error function defined as:
\begin{equation} \partial_{x}\left(Erf\left[ \frac{\sqrt{2}(x)}{b}\right] \right) =\frac{2}{b}\sqrt{\frac{2}{\pi}}\exp^{-\frac{2x^2}{b^2}}.
\end{equation}
In the limit case considered, where $W/b>>1$, the error function can be approximated by $Erf[(-W/\sqrt{2}b)]\approx -1$. Moreover, the first term in Eq. \ref{eq:lambda} has a small contribution in this limit because of its exponential dependence, and then $\lambda (0)$ may be written as
\begin{equation}
\begin{split}
\lambda(0)= \frac{\beta}{16a_{cc}}
\left[b\sqrt{2\pi} \right]\,\,\,. 
\label{eq:lambda0}
\end{split}
\end{equation} 

The effect of the deformation on the electronic gaps are then obtained via the energy relation assuming $k_y=0$, and the energy dependence for the n-th band is given by
\begin{equation}
\begin{split}
\epsilon_n=\hbar v_f\left( s \frac{2\lambda(0)}{W}\alpha +\left|\frac{n\pi}{W}- K\right| \right) \,\, ,
\end{split}
\label{eq:energy}
\end{equation}
with s = +1 when $({n\pi}/{W}-K)< 0$ and s=-1, otherwise.

Notice that: (i) $\epsilon_n$ has a linear dependence on $\alpha$, and fixed $b$, as shown in the numerical results presented in Fig. \ref{fig:FigGapVsAlpha} (a); (ii) in the case of the $3m+1$ armchair nanoribbon family, a band crossing is observed. There is a change of the band index value $n$ that corresponds to the minimum energy value. This leads to a slope change in the linear dependence of the gap size on $\alpha$, as shown in the Fig. \ref{fig:FigGapVsAlpha} (left panel); (iii) the maximum gap value ($\epsilon_g$)  can be evaluated obtaining the $\alpha$ value where the lowest two bands cross,  $\alpha_g=[\pi(n+1/2)-KW]/2\lambda(0)$, with the same value for different $3m+1$ armchair nanoribbons. For this case $\epsilon_g=\pi\hbar v_f/2W$, and therefore, the maximum gap value depends only on the nanoribbon width (see Fig. \ref{fig:FigGapVsAlpha}(b)); (iv) the additional dependence of the gap size on $\alpha$, but for a fixed $A$ value and varying $b$ width, shown in Fig.\ \ref{fig:FigGapVsAlpha}(c), may be derived from Eq. 12,  showing the radical dependence found in the numerical calculation; (v) Similar results can be obtained for the other families, but band crossings do not take place in the $3m$ case, and the gap evolution is fully described by Eq. \ref{eq:energy} with a single $n$ band. The lowest energy band decreases in energy, reaching null gap, and increases in energy according with the deformation considered.

The strained ribbon width $W_S$ may be calculated considering the infinitesimal element in $x$ direction which is modified by a scaling factor, $dx'=(1+\varepsilon_{xx})dx$, given by
\begin{equation}
\begin{split}
W_S=\int^W_0  (1+\varepsilon_{xx})dx \approx W + \frac{A^2 }{b}\frac{\sqrt{2\pi}}{4}\,\, ,
\end{split}
\label{eq:arclength}
\end{equation}

Notice that we have once more used the limit $W/b>>1$ to approximate the integration. In this way and using Eq. \ref{eq:lambda0}, the width variation is $\Delta W = \alpha \lambda(0)4 a_{cc}/\beta$. The energy for each single band is then simplified written as 
\begin{equation}
\begin{split}
\epsilon_n=\hbar v_f\left( \frac{s \beta}{2 a_{cc}} \frac{\Delta W}{W} +\left|\frac{n\pi}{W}- K\right| \right) \,\, .
\end{split}
\end{equation}
Therefore the gap value depends linearly on the ribbon percent differential length $\Delta W (\%)$, as predicted in the numerical results shown in Fig. \ref{fig:FigGapVsDeltaL}, with an angular coefficient independent of the deformation parameters A and b.

\subsection{Metallic AGNRs: unstrained and strained ribbons}
\begin{center}
\begin{figure}[h!]
\includegraphics[width=1.0\columnwidth]{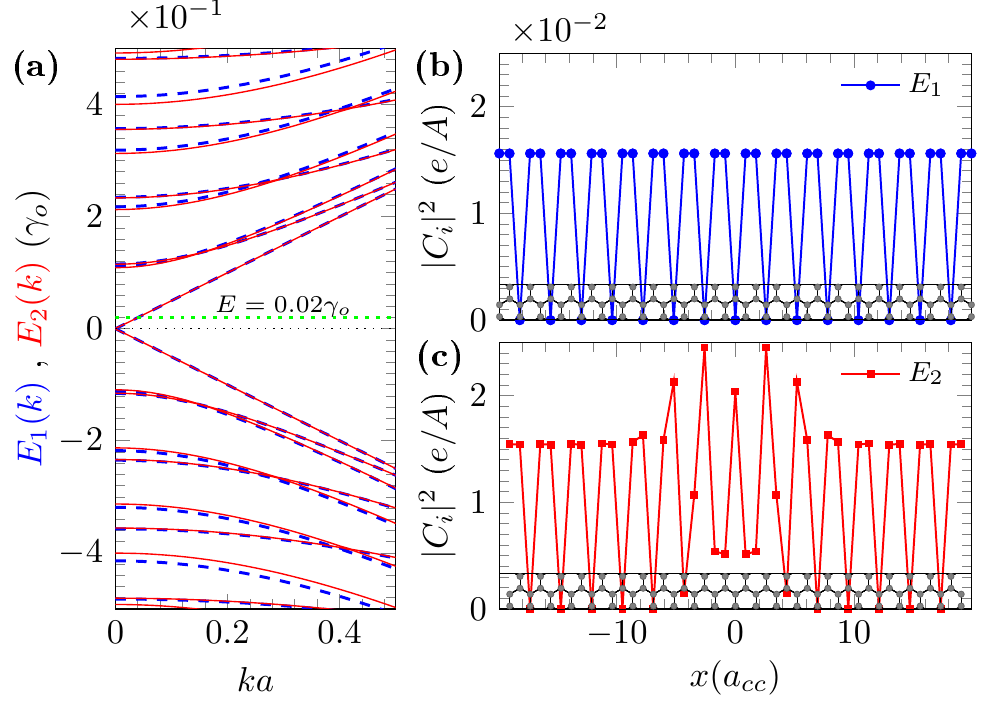}
\caption{(Color online) (a) Comparison between the electronic band structures of a metallic 47-AGNR (dashed blue curves) and a semiconducting 45-AGNR  (continuous red curves) with a mechanical fold deformation in which the gap is closed ($\alpha=18.9\%$). Electronic probability density for the (b) metallic 47-AGNR  and for the (c) semiconducting 45-AGNR under the mechanical deformation $\alpha=18,9\%$, both at the energy $0.02 eV$, marked with green dotted line in (a). The unitary cells for each AGNR are show at the bottom of (b) and (c).} 
\label{fig:metallic45-47}
\end{figure} 
\end{center}

\begin{center}
\begin{figure}[h!]
\includegraphics[width=1.05\columnwidth]{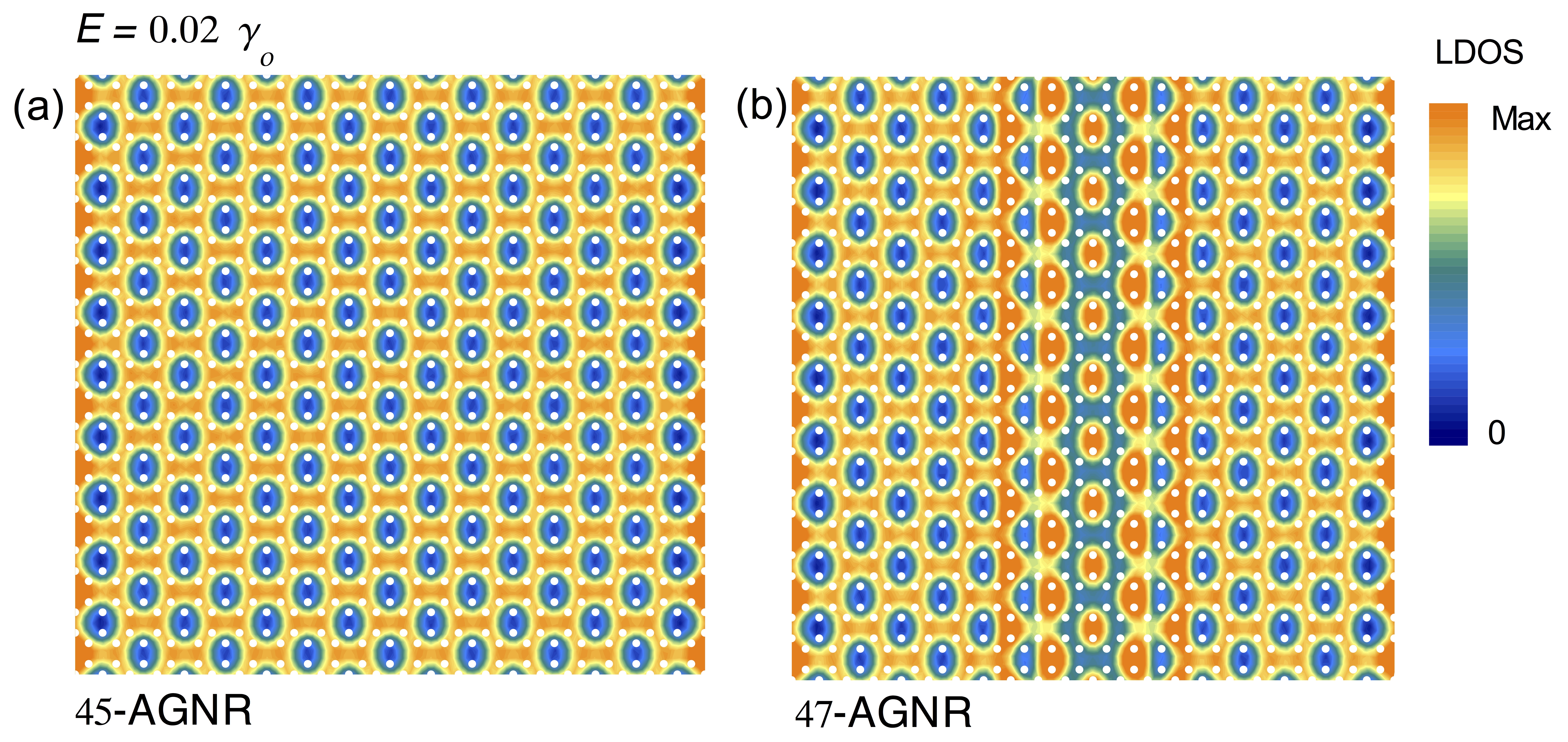}
\caption{(Color online) Contour plot of the Local electronic density of states for the (a) metallic 47-AGNR  and the (b) semiconducting 45-AGNR under the mechanical deformation $\alpha=18,9\%$, at the energy $0.02 \gamma_0$. White dots represent the atomic positions in the unstrained lattice.} 
\label{fig:LDOSmetallic}
\end{figure} 
\end{center}

We have also investigated the main differences between an unstrained metallic AGNR  and an original semiconducting ribbon that under a particular strain closes the conductance gap and behaves like a metallic system. This is the case of the $45-AGNR$ that under the mechanical deformation given by $\alpha=18.9\%$ ($b=5.5\,a_{cc}$) exhibits null gap. Both electronic band structures are shown in Fig. \ref{fig:metallic45-47}(a): the similarity between the electronic bands are remarkable in the energy range close to the Fermi energy. Substantial differences are found, however, in the electronic probability density, given by the electronic wave function coefficients $|c_{i}|^{2}$. The electronic probability densities for both unstrained and strained metallic AGNRs, are shown in Fig.\ref{fig:metallic45-47}(b) and (c), respectively, at the energy $E=0.02\gamma_0$. The strain considered in the 45-AGNR induces a probability enhancement at the center of the ribbon for all energies investigated in the first band. These findings are evident in the results of the local density of states at the same energy ($E=0.02\gamma_0$) depicted in Fig. \ref{fig:LDOSmetallic} (a) and (b), for the 47-AGNR and 45-AGNR, respectively, and for the same strain parameters considered in the previous figure. The electronic concentration of charge at the central part of the fold deformation may allow, for instance, better conditions for functionalization of the ribbon. To go further in this direction, a self-consistent calculation is sometimes required \cite{Barraza2005,Gibertini2010} to properly take into account the probable redistribution of the $\pi$-electrons caused by the deformation and the foreign molecule that may induce an electrostatic Coulomb potential. In the present case of armchair nanoribbons, such self-consistent calculation is not essential since we are considering soft mechanical deformations that do not induce robust charge redistribution.

\section{Conclusions}

In summary, possible electro-mechanical applications could be engineered by straining AGNRs, eventually turning them into metallic or semiconducting ribbons by tuning with appropriate folding parameters. Numerical results for the conductance gap were derived by following recursive Green's function protocols. The numerical results were well explained by following an analytical description within the Dirac equation. We have found that the energy range of the band gaps are well bounded by the deformation parameters. Also, the results show that the same maximum bandgap is reached by the 3m+1 AGNR family at a certain deformation, at different amplitudes for different but fixed fold widths, and that the maximum bandgap is defined by the AGNRs width. 

By using the same system, a bounded bandgap can be tuned just by being strained as a folded ribbon. Then the desired gap energy may be obtained by appropriately adjusting the strained-fold parameter. On the other hand, the semiconducting 3m family turns out metallic for a particular deformation, independently of the AGNRs undeformed width. These findings are summed out in an universal behavior for the bandgap with respect to a percent width difference, verified in both tight binding and continuum model descriptions. Finally, even though AGNRs cannot be used as electronic waveguide as strained fold ZGNRs due to the null induced pseudomagnetic fields, it may alternatively be applied as sensor devices as it opens the possibility for better functionalization scenarios.

\section{Acknowledgments}
We thank N. Sandler and R. Carrillo-Bastos for interesting discussions. This work has been financially supported by FAPERJ under grant E-26/102.272/2013. We acknowledge the financial support from CNPq and from the INCT de Nanomateriais de carbono.


\begin{thebibliography}{42}

\bibitem{Castro2009} A. H. Castro Neto, F Guinea, N. M. R. Peres, K. S. Novoselov, and A. K. Geim, Rev. Mod. Phys. 81, 109 (2009).

\bibitem{Wakabayashy2010}  K. Wakabayashi, K-I Sasaki, T. Nakanishi and T. Enoki, Science and Technology of Advanced Materials 11, 5 (2010).

\bibitem{Ritter2008} C. Ritter, S. S. Makler, and A. Latg\' e, Phys. Rev. B 77, 195443 (2008).

\bibitem{Brey2006} L. Brey and H. A. Fertig, Phys. Rev. B 73, 235411 (2006).
 
\bibitem{Mucciolo2009} E. R. Mucciolo, A. H. Castro Neto, and  C. H. Lewenkopf, Phys. Rev. B 79, 075407 (2009).

\bibitem{Levy2010} N. Levy, S. A. Burke, K. L. Meaker, M. Panlasigui, A. Zettl, F. Guinea, A. H. C. Neto, and M. F. Crommie, Science 329, 544 (2010).

\bibitem{Lim2015} H. Lim, J. Jung, R. S. Ruo, and Y. Kim, Nat. Commun. 6, 9601 (2015).

\bibitem{Downs2016} C. S. C. Downs, A. Usher, and J. Martin, J. Appl. Phys. 119, 194305 (2016).

\bibitem{Pereirax2009} V. M. Pereira and A. H. Castro Neto, Phys. Rev. Lett. 103, 046801 (2009).

\bibitem{Klimov2012} N. N. Klimov, S. Jung, S. Zhu, T. Li, C. A. Wright, S. D. Solares, D. B. Newell, N. B. Zhitenev, and J. A. Stroscio, Science 336, 1557 (2012).

\bibitem{Li2015} S.-Y. Li, K.-K. Bai, L.-J. Yin, J.-B. Qiao, W.-X. Wang, L. He, Phys. Rev. B 92, 245302 (2015).

\bibitem{Guinea2010} F. Guinea, M. I. Katsnelson, and A. K. Geim, Nat. Phys. 6, 30 (2010).

\bibitem{Bao2009} W. Bao, F. Miao, Z. Chen, H. Zhang, W. Jang, C. Dames, and C. N. Lau, Nat. Nanotech. 4, 562 (2009).

\bibitem {McEuen} J. S. Bunch, S. S. Verbridge, J. S. Alden, A. M. van der Zande, J. M. Parpia, H. G. Craighead, and P. L. McEuen, Nano Lett. 8, 2458 (2009).

\bibitem {Fogler} M. M. Fogler, F. Guinea, and M. I. Katsnelson, Phys. Rev. Lett. 101, 226804 (2008).

\bibitem{Schneider2015} M. Schneider, D. Faria, S. Viola Kusminskiy, and N. Sandler, Phys. Rev. B 91, 161407(R) (2015).

\bibitem{Wakker2011} G. M. M.Wakker, R. P. Tiwari, and M. Blaauboer, Phys. Rev. B 84, 195427 (2011).

\bibitem{Pereira2009} V. M. Pereira, A. H. Castro Neto, and N. M. R. Peres, Phys. Rev. B 80, 045401 (2009).

\bibitem{Moldovan2013} D. Moldovan, M. Ramezani Masir, and F. M. Peeters, Phys. Rev. B 88 035446 (2013). 

\bibitem{Ramon2014} R. Carrillo-Bastos, D. Faria, A. Latg\' e, F. Mireles, and N. Sandler, Phys. Rev. B 90, 041411(R) (2014).

\bibitem {Daiara2013} D. Faria, A. Latg\' e, S. E. Ulloa, and N. Sandler, Phys. Rev. B 87, 241403 (2013).

\bibitem{Zenan2013} Zenan Qi, D. A. Bahamon, Vitor M. Pereira, Harold S. Park, D. K. Campbell, and A. H. Castro Neto, Nano Lett. 13, 2692 (2013).

\bibitem{Ramon2016}  R. Carrillo-Bastos, C. Leon, D. Faria, A. Latg\' e, E. Y. Andrei, and N. Sandler, Phys. Rev. B 94, 125422 (2016).

\bibitem{Polini2011} D. Rainis, F. Taddei, M. Polini, G. Leon, F. Guinea, and V. I. Fal’ko, Phys. Rev. B 83, 165403 (2011).

\bibitem{Xu2012} P. Xu, Y. Yang, S. D. Barber, M. L. Ackerman, J. K. Schoelz, D. Qi, I. A. Kornev, L. Dong, L. Bellaiche, S. Barraza-Lopez, and P. M. Thibado, Phys. Rev. B 85, 121406 (2012).


\bibitem{Prada2010} E. Prada, P. San-Jose, and L. Brey, Phys. Rev. Lett. 105, 106802 (2010).

\bibitem{Lu2010} Y. Lu and J. Guo, Nano Res. 3, 189, (2010).

\bibitem {Sun2008} L. Sun, Q. Li, H. Ren, Q. W. Shi, and J. Yang, The J. of Chem. Phys. 129, 074704 (2008).

\bibitem{Landau1970} L. Landau and E. M. Lifshitz, {\it{Theory of Elasticity (Volumen 7 of A Course of Theoretical Physics)}}
(Pergamon Press, Cambridge, 1970).

\bibitem{Katesnelson2012} M. I. Katesnelson, {\it{Graphene: Carbon in Two Dimensions}} (Cambridge University Press, 2012).

\bibitem{Castro2014} Z. Qi, A. L. Kitt, H. S. Park, V. M. Pereira, D. K. Campbell, and A. H. Castro Neto, Phys. Rev. B 90, 125419 (2014).

\bibitem{Sloan2013} J. V. Sloan, Alejandro A. Pacheco Sanjuan, Z. Wang, C. Horvath, and S. Barraza-Lopez, Phys. Rev. B 87, 155436  (2013).

\bibitem{Settnes2016} M. Settnes, S. R. Power, and A.-P. Jauho, Phys. Rev. B 93, 035456 (2016).

\bibitem {Ando} H. Suzuura and T. Ando, Phys. Rev. B 65, 235412 (2002).

\bibitem{Vozmediano2010} M. Vozmediano, M. Katsnelson, and F. Guinea, Physics Reports 496, 109 (2010).

 \bibitem{Barraza89} A. A. Pacheco Sanjuan, Z. Wang, H. P. Imani, M. Vanevi\'c, and S. Barraza-Lopez, Phys. Rev. B 89, 121403(R) (2014).

\bibitem{Datta} S. Datta, {\it{ Electronic Transport in Mesoscopic System}}(Cambridge: Cambridge University Press, 1997).

\bibitem{Latge2000}	M. S. Ferreira, T. G. Dargam, R. B. Muniz, and A. Latg\'e Phys. Rev. B 62, 16040 (2000).

\bibitem{Caio2013} C. H. Lewenkopf and E. R. Mucciolo, Jour. Comp. Electronics 12, 203 (2013).

\bibitem{Thorgisson2014} G. Thorgisson et al, J. Comp. Phys. 261, 256 (2014).

\bibitem{Barraza2005} S. Barraza-Lopez, S. V. Rotkin, Y. Li, and K. Hess,  Europhys. Lett. 69, 1003 (2005).

\bibitem{Gibertini2010} M. Gibertini, A. Tomadin, M. Polini, A. Fasolino, and M. I. Katsnelson,
Phys. Rev. B 81, 125437  (2010).


\end{thebibliography}
\end{document}